\documentclass[11pt,xcolor=dvipsnames,fleqn]{article}
\DeclareMathAlphabet{\scr}{U}{rsfs}{m}{n}

\pdfoutput=1
\usepackage{scalerel,stackengine}
\usepackage{subfigure}
\usepackage{latexsym}
\usepackage{epsfig}
\usepackage[mathscr]{eucal}
\usepackage{amsfonts}
\usepackage{amscd}
\usepackage{multirow}
\usepackage{cite}
\usepackage{array}
\usepackage{amssymb}
\usepackage{amsthm}
\usepackage[outdir=./]{epstopdf}
\usepackage{colordvi}
\usepackage[centertags]{amsmath}
\usepackage{indentfirst}
\usepackage{geometry}
\usepackage{float} 
\usepackage{enumerate}
\usepackage{graphicx}
\usepackage{booktabs}
\usepackage{theorem}
\usepackage[footnotesize]{caption}
\usepackage{soul}
\usepackage{mcite}
\usepackage{slashed}%slash notation for gamma matrices
\usepackage{braket}%for bra and ket notations
\usepackage{color,xcolor}
\usepackage{bbm}

\usepackage[utf8]{inputenc}
\usepackage{fancyvrb}% provides improved Verbatim command
\usepackage{framed}
\usepackage{xspace}
\usepackage{todonotes}
\usepackage{siunitx}%Allows for formatting units via \SI{number}{Unit}

\usepackage{psfrag,rotating}
\usepackage{epstopdf}
\usepackage{epsf,graphics}
\usepackage{nccmath}
\numberwithin{figure}{section}
\numberwithin{table}{section}
\def\zz{\mathbb{Z}}

\allowdisplaybreaks

\usepackage{cleveref}
\crefname{chapter}{Chapter}{Chapter}
\crefname{section}{Sec.}{Secs.}
\crefname{table}{Tab.}{Tabs.}
\crefname{figure}{Fig.}{Figs.}
\crefname{equation}{Eq.}{Eqs.}
\crefname{appendix}{Appendix\ }{Appendix\ }

\begin{document}
\pdfoutput=1

\title{
    \vspace*{-3.7cm}
    \vspace*{2.7cm}
    \textbf{Revisiting one-loop corrections to the trilinear Higgs boson self-coupling in the Inert Doublet Model \\[4mm]}}

\date{}
\author{
Jaouad El Falaki$^{1\,}$\footnote{E-mail: \texttt{jaouad.elfalaki@gmail.com,j.elfalaki@uiz.ac.ma}}
\\[5mm]
{\small\it $^1$ LPTHE, Physics Department, Faculty of Sciences, Ibnou Zohr University, P.O.B. 8106 Agadir, Morocco.} \\[7mm]
\textit{I would like to dedicate this paper to my sister Fatima, who is bravely fighting against cancer}
}

\maketitle

\begin{abstract}
We investigate predictions of the trilinear Higgs self-coupling with radiative corrections in the context of the Inert Doublet Model. The triple Higgs vertex is computed at the one-loop level based on the on-shell renormalization scheme. We calculate its possible deviation from the predictions within the standard model, taking into account all relevant theoretical and experimental constraints, including  dark matter searches and the latest bounds on the branching fraction of the Higgs boson decaying to invisible particles. By scanning the model's  parameter space, we find that the deviation in the triple Higgs boson self-coupling from standard model expectations can be substantial, exceeding 100\% in certain regions of the parameter space.
\end{abstract}
\thispagestyle{empty}
\vfill
\newpage
\setcounter{page}{1}
%=================================================================================
\section{Introduction}
\label{sec:introduction}
%%%%%%%%%%%%%%%%%%%%%%%%
A great achievement in the history of high energy physics was made on July 4, 2012, with the discovery of the Higgs boson by ATLAS and CMS, at the CERN Large Hadron Collider (LHC)\cite{Aad:2012tfa,Chatrchyan:2012xdj}. Since then, the spectrum of the Standard Model (SM) of particle physics has been completed. Following this discovery,  more accumulated data has been analyzed during LHC Run I-II and it has been found that the properties of the Higgs boson are in perfect agreement with the  predictions from the SM, with a level of accuracy of $10-20\%$\cite{ATLAS:2019nkf,CMS:2018uag}.\\
To establish experimentally the Brout-Englert-Higgs mechanism of electroweak symmetry breaking (EWSB), it is necessary to measure not only the couplings of the Higgs boson with the fermions and the gauge bosons, but also the self-coupling of the Higgs boson, i.e. the triple  $\lambda_{hhh}$ and the quartic $\lambda_{hhhh}$ Higgs boson couplings, in order to be able to reconstruct the shape of the  Higgs scalar  potential. The measurement of the Higgs boson quartic coupling is more challenging because the cross section for triple Higgs production at the LHC is much smaller (the cross section for $pp\rightarrow 3h$ production is about 0.1 fb at $\sqrt{s}= 14 \,TeV$), and it is out of reach even for the high-luminosity LHC (HL-LHC)\cite{Cepeda:2019klc}. On the other hand, the trilinear self-coupling of the Higgs boson $\lambda_{hhh}$ can be measured from the production of a pair of Higgs bosons at the LHC \cite{Abouabid:2021yvw}. 
Recently, a statistical combination by ATLAS has been presented \cite{ATLAS:2022kbf}, in which the bound on  $\lambda_{hhh}$  has been significantly reduced to $ -0.4<\kappa_\lambda<6.3$ at $95\%$ confidence level, where  $\kappa_\lambda \equiv \frac{\lambda_{hhh}}{\lambda_{hhh}^{SM}}$ is the self-coupling modifier parameter. CMS also has derived a limit on $\kappa_\lambda$ which is constrained to be within $ -1.24<\kappa_\lambda<6.49$  at  $95\%$ CL \cite{CMS:2022dwd}. The measurement accuracy of $\lambda_{hhh}$ will be improved at future experiments such as the HL-LHC  where $\kappa_\lambda$ is constrained to be between 0.1 and 2.3  at $95\%$ CL with 3 $ab^{-1}$ data\cite{Cepeda:2019klc}. At future linear collider  such as the ILC, the triple Higgs boson coupling is expected to be measured at the precision level of $27\%$  in the double Higgs-strahlung process $e^+e^- \rightarrow Zhh$ at $\sqrt{s}=500\, GeV$ with an integrated luminosity of 4 $ab^{-1}$\cite{Fujii:2020pxe,ILCInternationalDevelopmentTeam:2022izu}. A relative precision of $10\%$  on $\lambda_{hhh}$   is also possible at $1\,TeV$ from the di-Higgs production in WW fusion process $e^+e^- \rightarrow \nu \bar{\nu} hh$ with an accumulated 8 $ab^{-1}$ of integrated luminosity \cite{Fujii:2020pxe,ILCInternationalDevelopmentTeam:2022izu}. The  expectation for these precise measurements motivates the study of the radiative corrections to $\lambda_{hhh}$.\\
The SM is unable to explain certain phenomena such as dark matter, the hierarchy problem and tiny neutrino masses. As a result, new physics beyond the SM (BSM) is needed to provide answers to these unsolved problems. The Higgs sector of the SM  only includes one Higgs doublet, but there is no fundamental reason to believe it must be minimal. Among popular BSM candidates are models with extended Higgs sectors, such as the Inert Doublet Model (IDM).
 Originally proposed by Deshpande and  Ma \cite{Deshpande:1977rw} and  initially suggested for EWSB studies, this model is highly intriguing due to its potential to generate tiny neutrino masses \cite{Ma:2006km}, provide a dark matter candidate \cite{Gustafsson:2007pc,Hambye:2007vf,Agrawal:2008xz,Dolle:2009fn,Cao:2007rm,LopezHonorez:2006gr,Goudelis:2013uca}, and address the naturalness problem \cite{Barbieri:2006dq}. 
 The IDM consists of adding a second Higgs doublet that does not acquire a vacuum expectation value (VEV) and has no coupling with SM fermions.  An exact $\mathbb{Z}_2$ symmetry is imposed, with the SM Higgs doublet being even and the additional scalar doublet being odd. The preserved $\zz_2$ symmetry ensures that the extra doublet  does not interact with matter and its lightest stable neutral component can act as a dark matter particle. After EWSB, the IDM has a spectrum of five physical scalars: a CP-even Higgs boson $h$ (identified with the discovered SM Higgs), and four inert scalars ($H$, $A$, and $H^\pm$).
The rich phenomenology of the IDM  has been thoroughly studied in the literature, both in the context of future Higgs factories like the ILC and CLIC and at the LHC \cite{Dolle:2009ft,Aoki:2013lhm,Arhrib:2012ia, Krawczyk:2013jta,Arhrib:2014pva, Datta:2016nfz,Dutta:2017lny,Kalinowski:2018ylg,Kalinowski:2018kdn,Guo-He:2020nok,Yang:2021hcu,Kalinowski:2020rmb,Melfo:2011ie,Abercrombie:2015wmb,Ilnicka:2015jba,Blinov:2015qva,Poulose:2016lvz,Hashemi:2016wup,Wan:2018eaz,Belyaev:2018ext,Dercks:2018wch,Lu:2019lok,Gil:2012ya,Swiezewska:2015paa,Blinov:2015vma,Huang:2017rzf,Belyaev:2016lok,Arhrib:2013ela,Eiteneuer:2017hoh,Ghosh:2021noq,Jueid:2020rek,Krawczyk:2013pea,Enberg:2013ara,Krawczyk:2013wya,Kanemura:2018esc,Hashemi:2015swh,Ahriche:2018ger}. Additionally, there have been many studies of the model that go beyond the lowest order of the perturbation, including those at one-loop level \cite{Arhrib:2015hoa,Abouabid:2022rnd,Kanemura:2016sos,Abouabid:2020eik, Aiko:2021nkb, Ramsey-Musolf:2021ldh,Banerjee:2019luv,Banerjee:2021oxc,Banerjee:2021anv,Banerjee:2021xdp,Banerjee:2021hal,Banerjee:2016vrp,Kanemura:2019kjg,Kanemura:2019slf,Kanemura:2017gbi,Ferreira:2015pfi} and two-loop order \cite{Braathen:2019zoh,Braathen:2019pxr, Senaha:2018xek}.
The one-loop contributions to the triple Higgs boson coupling from Standard Model particles have been investigated in \cite{Kanemura:2004mg,Kanemura:2002vm,Kanemura:2015mxa,Hollik:2001px,Arhrib:2015hoa}, where it was found that these corrections are dominated by loops involving the top quark.  
The radiative corrections to $\lambda_{hhh}$ in some non-supersymmetric Higgs models can be found in references \cite{Kanemura:2004mg,Kanemura:2002vm,Bahl:2022jnx,Braathen:2020vwo,Kanemura:2017wtm,Kanemura:2016lkz,He:2016sqr,Kanemura:2015mxa,Aoki:2012jj,Basler:2017uxn,Basler:2019iuu,Moyotl:2016fdk,Krause:2016oke,Osland:2008aw}, and for corrections in certain supersymmetric models, see for example Refs 
\cite{Barger:1991ed,Hollik:2001px,Borschensky:2022pfc,Brucherseifer:2013qva,Nhung:2013lpa,Dobado:2002jz,Kanemura:2010pa}.
These new physics effects in models with extended Higgs sectors have been shown to be large and can significantly enhance the $\lambda_{hhh}$ coupling in a wide range of parameter space. Calculations of Higgs boson couplings that include higher-order corrections are mandatory to compare theory predictions with future precision data from hadron and lepton colliders. In Ref \cite{Arhrib:2015hoa},  the one-loop contributions of the inert scalars  to $\lambda_{hhh}$  are discussed only in the  degenerate spectra, i.e., $m_{H}=m_{A}=m_{H^\pm}$. In addition, one loop corrections to $\lambda_{hhh}$ within the IDM have also been discussed in some scenarios \cite{Kanemura:2016sos}, but under the assumption that $m_{H}=m_{H^\pm}$.
In the present letter, we will compute the radiative corrections to the triple Higgs self-coupling considering non-degenerate masses for the inert scalars, while taking into account all current theoretical and experimental constraints on the  IDM.\\
The layout of the letter is as follows: In Sec.~\ref{idms}, we  briefly introduce the 
IDM and outline its theoretical and experimental constraints. In Sec.~\ref{sec::3}, we present the on-shell renormalization scheme and provide a comprehensive explanation of the triple Higgs coupling at the one-loop level. In Sec.~\ref{sec::4}, we present our numerical results for the SM and IDM. Conclusions are given
in the last section.
%%%%%%%%%%%%%%%%%%%%%%%%%%%%%% The IDM %%%%%%%%%%%%%%%%%%%%%%%%%%

\section{The Inert Doublet Model}
\label{idms}
\subsection{The Model}
The IDM is a simple extension of the Standard Model of particle physics which consists of the SM, including its Higgs doublet $\Phi_1$, and 
an additional  $SU(2)$ doublet $\Phi_2$ that brings in four new scalar particles. An exact $\zz_2$ symmetry is introduced such that the SM doublet is even $\Phi_1\longrightarrow \Phi_1 $ while the added extra doublet (inert doublet) is odd $\Phi_2\longrightarrow -\Phi_2$. This unbroken $\zz_2$ parity guarantees the absence of coupling between fermions and the inert doublet $\Phi_2$, therefore there is no flavor-changing neutral currents.  Moreover, it ensures that the lightest neutral component of $\Phi_2$ is a natural dark matter candidate.
The decomposition for the two doublets around the vacuum state is given by:
\begin{eqnarray}
	\Phi_1 = \left (\begin{array}{c}
		G^\pm \\
		\frac{1}{\sqrt{2}}(v + h + i G^0) \\
	\end{array} \right)
	\qquad , \qquad
	\Phi_2 = \left( \begin{array}{c}
		H^\pm\\ 
		\frac{1}{\sqrt{2}}(H + i A) \\ 
	\end{array} \right) 
\end{eqnarray}
Where only the SM doublet $\Phi_1$ is involved in  EWSB, $G^0$ and $G^\pm$ correspond to the three Nambu-Goldstone bosons gauged away by  the longitudinal component of $Z$ and $W^\pm$ respectively, $h$ is the SM Higgs boson and $v$ is the VEV of the SM Higgs doublet. The second doublet $\Phi_2$ does not participate in EWSB and it contains four new inert scalars $H$, $A$ and $H^\pm$ where either $A$ or $H$ may act as potential dark matter candidate, depending on the mass hierarchy of these two inert scalars.\\
The most general renormalizable tree-level scalar potential in this model can be written as:
\begin{eqnarray}
	V  &=&  \mu_1^2 |\Phi_1|^2 + \mu_2^2 |\Phi_2|^2  + \lambda_1 |\Phi_1|^4
	+ \lambda_2 |\Phi_2|^4 +  \lambda_3 |\Phi_1|^2 |\Phi_2|^2 + \lambda_4
	|\Phi_1^\dagger \Phi_2|^2 \nonumber \\
	&+&\frac{\lambda_5}{2} \left\{ (\Phi_1^\dagger \Phi_2)^2 + {\rm h.c} \right\},
	\label{idmpotential}
\end{eqnarray}
where $\mu_1$ and $ \mu_2$ are the mass of the $\Phi_1$ and $\Phi_2$ fields, and all $\lambda_{1,2,3,4}$  parameters are real since the above potential must be hermitian while the phase of $\lambda_5$ can be absorbed into an appropriate redefinition of $\Phi_1$ and $\Phi_2$ fields.\\ After EWSB the five scalar particles of the model acquire their masses which are given by:
\begin{eqnarray}
	&& m_{h}^2 = - 2 \mu_1^2 = 2 \lambda_1 v^2 \nonumber \\
	&& m_{H}^2 = \mu_2^2 + \lambda_L v^2 \nonumber \\
	&&  m_{A}^2 = \mu_2^2 + \lambda_S v^2 \nonumber \\
	&&  m_{H^{\pm}}^2 = \mu_2^2 + \frac{1}{2} \lambda_3 v^2
	\label{spect.IHDM}
\end{eqnarray}
where $\lambda_{L,S}$ are defined as:
\begin{eqnarray}
	\lambda_{L,S} &=& \frac{1}{2} (\lambda_3 + \lambda_4 \pm \lambda_5)
\end{eqnarray}
 The trilinear self-coupling of the Higgs boson at tree level in the IDM involves only SM parameters and is given by:
\begin{ceqn}
	\begin{align}
		\Gamma_{hhh}^{tree} = \frac{-3 m_{h}^2}{v} \label{hhhtree}
	\end{align}
\end{ceqn}
In the IDM, there are eight independent parameters: 5$\lambda_i$, 2$\mu_i$, and $v$.
After eliminating one parameter through the minimization condition and determining the VEV using the $W$ boson mass, we are left with six remaining independent parameters which will be selected as follows:
\begin{eqnarray}
	\{ \mu_2^2, \lambda_2, m_{h}, m_{H^\pm}, m_{H}, m_{A} \}
\end{eqnarray}
%===============================================
\subsection{Theoretical and Experimental Constraints}	
In this study, we explore the same parameter space as in our previous published paper~\cite{Abouabid:2020eik}. The IDM parameter space is obtained by performing an extensive parameter scan in the whole space with all experimental and theoretical constraints applied. The constraints used are summarized below:
	\begin{itemize}
		\item The theoretical constraints:
		\begin{itemize}
			\item The perturbative unitarity~\cite{Kanemura:1993hm,Akeroyd:2000wc}
			\item The vacuum stability~\cite{Branco:2011iw,Deshpande:1977rw}
			\item The inert vacuum and charge-breaking minima~\cite{Ginzburg:2010wa}
%			\item Inert Vacuum
%			\item Unitarity
		\end{itemize}
		\item The experimental constraints:
		\begin{itemize}	%
			\item The Higgs data from the LHC \cite{Aaboud:2018xdt,Sirunyan:2018ouh}. 
			\item The invisible Higgs decay \cite{ATLAS:2020kdi}.
			\item The direct collider searches at LEP \cite{Arhrib:2012ia,Swiezewska:2012eh,Belanger:2015kga,Lundstrom:2008ai}
			\item The electroweak precision tests \cite{Peskin:1991sw, Barbieri:2006dq, Tanabashi:2018oca}.
			\item The 
			dark matter searches \cite{Zyla:2020zbs, Abbott:1982af, Dine:1982ah, Preskill:1982cy,
				Belanger:2020gnr, Aprile:2018dbl, Amole:2019fdf, Abdelhameed:2019hmk, 
				Agnes:2018ves, Aaboud:2017phn, Sirunyan:2017hci, Belyaev:2018ext, Lu:2019lok}.
		\end{itemize}
	\end{itemize}

\section{Calculation of one-loop corrections to the triple Higgs coupling}
\label{sec::3}
In this section, we briefly discuss the renormalization of the trilinear self-coupling of the Higgs boson. Using the 't Hooft-Feynman gauge, we calculate the radiative corrections to the tree-level formula in Eq.~\ref{hhhtree} in both the SM and IDM including contributions from all particles in the loop. Figure~\ref{fig:fdiag} illustrates some of the Feynman diagrams contributing to the triple Higgs boson coupling. Note that the dimensional regularization has been used to evaluate the one-loop Feynman amplitudes.
The calculations are carried out using Mathematica packages FeynArts and FormCalc~\cite{Hahn:2000kx,Hahn:1998yk,Hahn:2006qw}. The numerical evaluation of the scalar one-loop integrals has been performed with the  LoopTools package~\cite{Hahn:1999mt,Hahn:2010zi}. It should be noted that the UV-finiteness of the virtual contributions has been cross-checked numerically and analytically. Compared to the general two Higgs doublet model (2HDM)\cite{Kanemura:2017wtm,Kanemura:2002vm,Kanemura:2015mxa}, the structure of the counter-terms for the triple Higgs boson coupling in the IDM is simpler and identical to those in the SM due to the absence of mixing between the SM Higgs boson and the inert scalars. We study an off-shell 
Higgs boson that decays into two real Higgs bosons $h^*(q)\to h(q_1)h(q_2)$ 
at the one-loop level, where $q$, $q_1$ and $q_2$ are the four-momenta of the three 
Higgs bosons satisfying an off-shell condition $q^2 \neq m_h^2$
for the decaying Higgs boson and on-shell conditions $q_1^2 = q_2^2 = m_h^2$ for the
two real Higgs bosons.
The UV divergences that emerge during the calculation's intermediate stages should eventually cancel out in the end. In order to do that, we adopt the on-shell renormalization scheme which is widely used in quantum field theory because it is simple to implement and it allows for a clear physical interpretation of the parameters of the theory. As in the SM, the tree-level trilinear Higgs self-coupling in Eq.~\ref{hhhtree} depends only on the VEV and Higgs boson mass. Hence, the renormalization procedure will be the same as the one adopted in the SM\cite{Bohm:1986rj,Hollik:1988ii,Denner:1991kt,Denner:2019vbn}. The SM fields and parameters are redefined as follows: 	
\begin{figure}[H]\centering
	\includegraphics[width=0.64\textwidth]{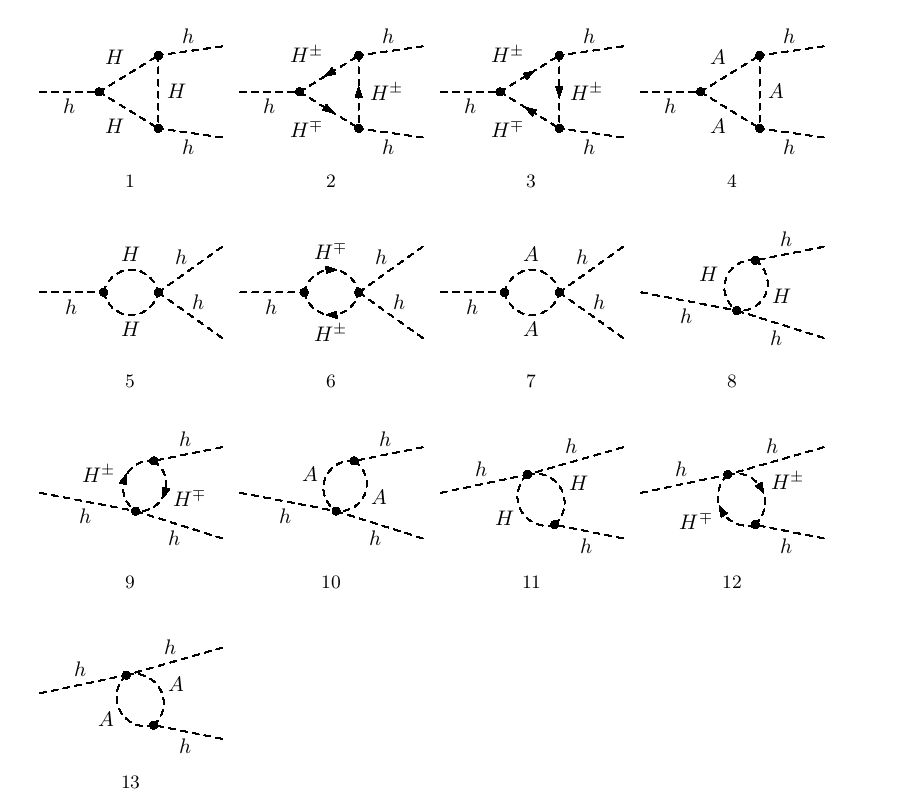}
	\caption{Some one-loop Feynman diagrams that do not exist in the SM contribute to the triple Higgs self-coupling within the IDM.
	}
	\label{fig:fdiag}
\end{figure}
\begin{eqnarray}
	&& m_h^2 \to m_h^2 + \delta m_h^2 \nonumber\\
	&& m_V^2 \to m_V^2 + \delta m_V^2\, , \, \quad V = Z, W^\pm \nonumber \\
	&& s_W \to s_W + \delta s_W \nonumber \\
	&& t \to t + \delta t \nonumber \\
	&& e \to (1 + \delta Z_e) e \nonumber\\ 
	&& Z^\mu \to \left(1 + \frac{1}{2} \delta Z_{ZZ}\right) Z^\mu + 
	\frac{1}{2} \delta Z_{ZA} A^\mu \nonumber\\
	&& A^\mu \to \left(1 + \frac{1}{2} \delta Z_{AA}\right) A^\mu + 
	\frac{1}{2} \delta Z_{AZ} Z^\mu \nonumber\\
	&& h \to Z_h^{1/2} h = \left(1 + \frac{1}{2} \delta Z_h \right) h 
\end{eqnarray}
where $s_W = \sin \theta_W$ is the Weinberg angle and 
$t = v (\mu_1^2 - \lambda_1 v^2)$ is the tadpole.  At tree level, the tadpole is zero if the minimization condition is satisfied, but it can receive finite corrections at the one-loop level. To ensure that the VEV of the Higgs field is consistent across all orders of perturbation theory, it is necessary to renormalize the Higgs tadpole. This can be done by adding a counter-term to the tadpole, which cancels out any divergences that appear at higher orders of perturbation theory. Consequently, we set the following condition: 
\begin{equation}
	\hat{T} =  \delta t + T =0\Longrightarrow  \delta t = -T  
\end{equation}
where $T$ is the one-loop contribution of 1PI diagrams.\\
The counter-terms of the masses are fixed by the following on-shell condition:
\begin{eqnarray}
	&& \textrm{Re} \hat{\Sigma}_{T}^{VV}(m_V^2) = 0 \Longrightarrow \delta m_V^2=\textrm{Re}\Sigma_{T}^{VV}(m_V^2) \quad V=W,Z \nonumber \\
	&& \textrm{Re} \hat{\Sigma}_{hh}(m_h^2) = 0 \Longrightarrow \delta m_h^2= \textrm{Re} \Sigma_{hh}(m_h^2)
\end{eqnarray}
where $\Sigma_{hh}$ and $\Sigma_{T}^{VV}$ are the one-loop non-renormalized self energies for the Higgs boson and gauge bosons respectively. 

By fixing the residue of the two-point Green functions to be equal to unity, the wave-function renormalization constant is determined from the following condition.
 \begin{eqnarray}
\textrm{Re} \frac{\partial \hat{\Sigma}_{hh} (k^2)}{\partial k^2} \Bigg|_{k^2 = m_h^2} =0 \Longrightarrow	\delta Z_{h} = -\textrm{Re} \frac{\partial {\Sigma}_{hh}}{\partial k^{2}}\Bigg|_{k^2=m_{h}^2}.
\end{eqnarray}
In the on-shell renormalization scheme, the electric charge is set by ensuring that there are no higher-order corrections to the $e^+e^- \gamma$ vertex in the Thomson limit. The electric charge renormalization constant  $\delta Z_e$ can be expressed as:
	\begin{equation}
	\delta Z_e=-\dfrac{1}{2}\delta Z_{AA}-\dfrac{s_W}{c_W}\dfrac{1}{2}\delta Z_{ZA}
\end{equation}
where
	\begin{equation}
	\delta Z_{AA}=-\dfrac{\partial\sum^{AA}_T( k^2)}{\partial k^2} \Bigg|_{k^2=0} and\quad \delta Z_{ZA}=2\dfrac{\sum^{AZ}_T(0)}{m_Z^2}
\end{equation}
To obtain the counter-term $\delta s_W$ one can use the on-shell definition of the weak mixing angle, which is defined as the ratio of the weak neutral current and the weak charged current. Thus, $\delta s_W$ is given by:
	\begin{equation}
		\delta s_W=\frac{c_W^2}{2s_W}\left(\frac{\delta m_Z^2}{m_Z^2} - \frac{\delta m_W^2}{m_W^2}\right)
\end{equation}	
By inserting the redefinition of the parameters into the Lagrangian, we obtain the following counter-term for the trilinear Higgs self-coupling.\\
\begin{eqnarray}
	&&\delta \Gamma_{hhh}= \frac{-3e^2}{2 s_W} \frac{m_h^2}{m_W} 
	\left(\delta Z_e - \frac{\delta s_W}{s_W} + \frac{\delta m_h^2}{m_h^2} + 
	\frac{e}{2 s_W} \frac{\delta t}{m_W m_h^2} 
	-\frac{\delta m_W^2}{2 m_W^2} + \frac{3}{2} \delta Z_h \right) \hspace{1.5cm}
	\label{counter-term}
\end{eqnarray}
To obtain the renormalized amplitudes for the triple Higgs coupling, the full one-loop one particle irreducible vertex $\Gamma_{hhh}^{1PI}(q^2, q_1^2, q_2^2)$ is added to the corresponding counter-terms $\delta \Gamma_{hhh}$ as follows:\\
\begin{eqnarray}
	&&\hat{\Gamma}_{hhh}(q^2, q_1^2, q_2^2)= \Gamma_{hhh}^{tree}+\Gamma_{hhh}^{1PI}(q^2, q_1^2, q_2^2)+
	\delta \Gamma_{hhh}
\end{eqnarray}

\section{Numerical Results}
\label{sec::4}
\begin{figure}[H]\centering
	\includegraphics[width=0.8\textwidth]{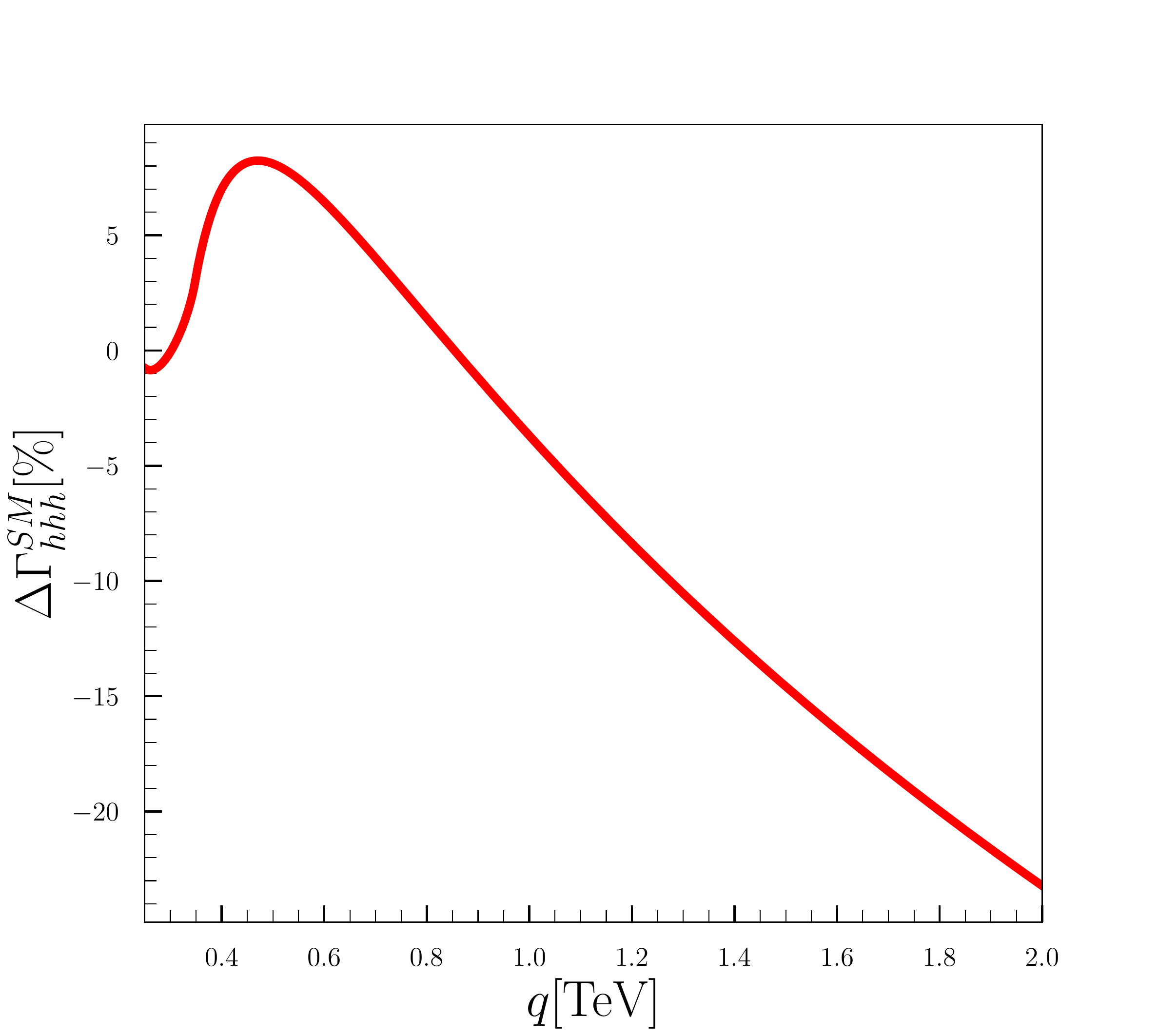}
	\caption{The relative correction $\Delta\Gamma_{hhh}^{SM}[\%]$ as a function of $q$, where $q^\mu$ is the four-momentum of the off-shell Higgs boson in $h^*\rightarrow hh$.}
	\label{fig:hhhsm}
\end{figure}

In this section, we present our numerical analysis for the triple Higgs coupling at the one-loop level in the SM and IDM. In order to parametrize the size of the radiative corrections and compare it to the Standard Model's predictions, we define the following ratio in the IDM:
\begin{eqnarray}
	&&\Delta \Gamma_{hhh}=\frac{\hat{\Gamma}_{hhh}(q^2, m_h^2, m_h^2)_{IDM}-\hat{\Gamma}_{hhh}(q^2, m_h^2, m_h^2)_{SM}}{\hat{\Gamma}_{hhh}(q^2, m_h^2, m_h^2)_{SM}}
\end{eqnarray}
While within the SM, we show our numerical results using the following relative ratio:
\begin{eqnarray}
	&&\Delta \Gamma_{hhh}^{SM}=\frac{\hat{\Gamma}_{hhh}(q^2, m_h^2, m_h^2)_{SM}-\Gamma_{hhh}^{tree}}{\Gamma_{hhh}^{tree}}
\end{eqnarray}
The following numerical values of the input parameters are adopted \cite{Tanabashi:2018oca}:
	\[ \begin{array}{lll}%
		m_{h}=125.18\, \text{GeV}  & m_{W}=80.379\, \text{GeV}  &  m_{t}=173.2\, \text{GeV}\\
		m_{\mu}=0.106\, \text{GeV}  &  m_{Z}=91.198\, \text{GeV} &  m_b=4.660\, \text{GeV}\\
		m_{\tau}=1.777\, \text{GeV} & \alpha=1/{137.036} \, &  m_c=1.275\, \text{GeV}
	\end{array}\]%
We scan the entire parameter space for the other IDM parameters, including physical masses $m_{A}$, $m_{H}$ and $m_{H\pm}$ as well as the $\mu_{2}^2$ parameter. We consider all relevant theoretical and experimental constraints and perform a random scan  over the IDM parameter space in the following ranges:
\begin{ceqn}
	\begin{align}		
	100 \textrm{ GeV} \leq m_{H} \leq 700 \textrm{ GeV} \nonumber \\
	20 \textrm{ GeV} \leq m_{A} \leq 62.5 \textrm{ GeV} \nonumber \\
	80 \textrm{ GeV} \leq m_{H^{\pm}} \leq 700 \textrm{ GeV} \nonumber \\
	0 \textrm{ GeV}^2 \leq \mu_2^2 \leq 10^6 
	\textrm{ GeV}^2
		\end{align}
	\end{ceqn}
It's worth mentioning that our numerical results are independent of $\lambda_{2}$ parameter. We will set $\lambda_{2}$  to a fixed value of 2.\\
It is noteworthy that in this letter, the inert scalar A is selected as the dark matter candidate. Furthermore, all points have been passed the upper bound from the invisible Higgs boson decay $Br(h\rightarrow AA)\le 11\%$\cite{ATLAS:2020kdi}.\footnote{It bears mentioning that the mass range between 20 GeV and 55 GeV of the dark matter candidate is ruled out from relic density constraints. It is also worth pointing out that the allowed values of $\mu_2^2$ range between 3000 $GeV^2$ and 4400 $GeV^2$ after passing all constraints.}\\
It is important to note that the mass splitting between the inert neutral scalars $H$ and $A$ is controlled by the $\lambda_5$ parameter, which ranges from 0.104 to 6.05 after all constraints have been taken into account. Additionally, $\lambda_5$ plays a crucial role in generating a tiny neutrino mass in the minimal scotogenic model, originally proposed by Ernest Ma in 2006 \cite{Ma:2006km}. This model provides an elegant way to generate tiny neutrino masses at the one-loop level and allows for the possibility of the dark matter particle being either the light Majorana fermion (Majorana dark matter) or the lightest among the $H$ and $A$ (scalar dark matter). The latter scenario, which involves scalar dark matter, is similar to the IDM. Thus, the permissible values of the $\lambda_5$ parameter in our analysis will be the same as those in the scotogenic model with scalar dark matter, where small values of the new Yukawa couplings that couple the dark doublet with the Majorana fermions generate tiny neutrino masses. However, in the case of Majorana dark matter, large values of the new Yukawa couplings are required to achieve the appropriate amount of dark matter relic density. In this scenario, the smallness of neutrino masses can only be achieved by ensuring a strong degeneracy between the two neutral scalars $H$ and $A$, leading to the suppression of the $\lambda_5$ parameter at $\lambda_5 \sim 10^{-9}$.\\
We visualize in Figure~\ref{fig:hhhsm} the size of the radiative corrections in the SM as a function of the four-momentum of the off-shell Higgs boson. The measurement of the triple Higgs coupling in future experiments will be done through the double Higgs production process. In this case, one of the Higgs bosons will be off-shell, which implies that the dependence on momentum $q$ is important. One can see that the total corrections to the trilinear self-coupling of the Higgs boson start from $-1\%$ around $q=250$ GeV and  can reach a maximum value of  $8.23\%$ for  $q=470$ GeV. It should be emphasized that the top-quark contribution is the dominant correction for large values of the momentum $q$.

\begin{figure}[H]\centering
	\includegraphics[width=0.62\textwidth]{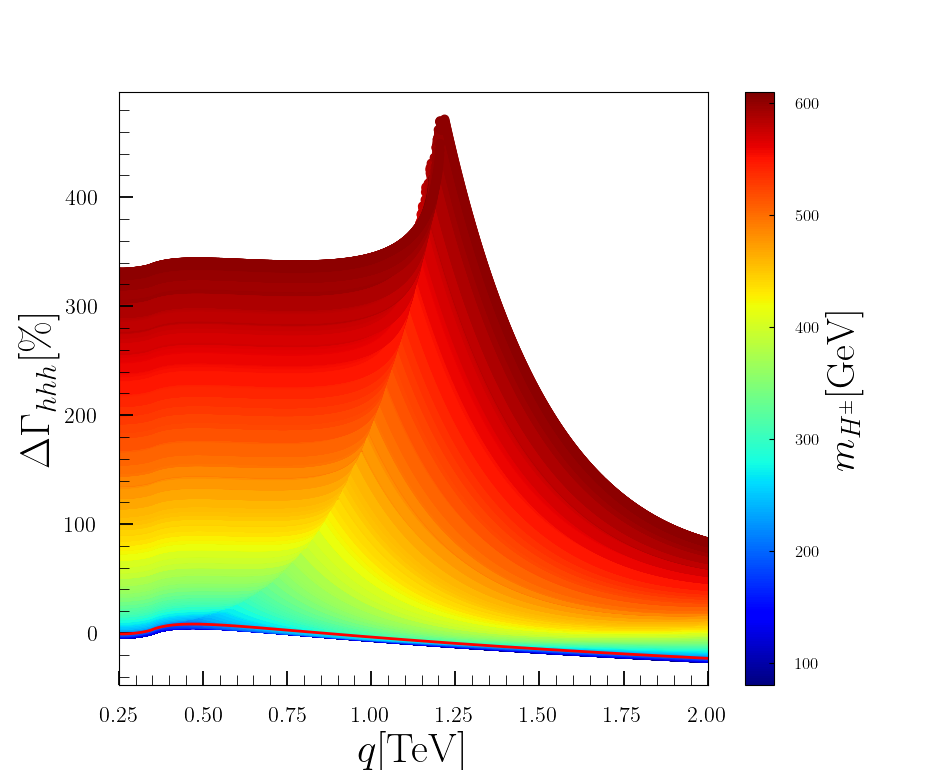}\includegraphics[width=0.62\textwidth]{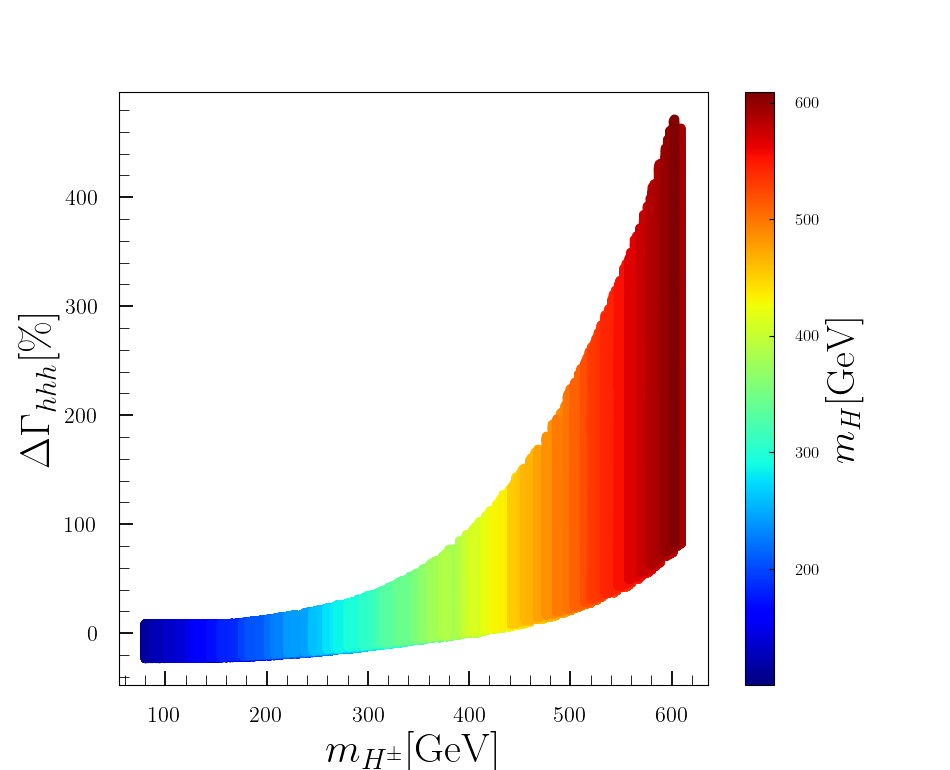}
	\caption{Left: $\Delta\Gamma_{hhh}$ as a function of the momentum $q$ where the the charged scalar mass $m_{H^{\pm}}$ is shown in the right column. Right: $\Delta\Gamma_{hhh}$ as a function of $m_{H^{\pm}}$ and the color code indicates the mass $m_H$ of the neutral scalar H. The red line in the left panel represents the relative ratio $\Delta\Gamma_{hhh}^{SM}$  in the SM.}
	\label{fig:hhhidm1}
\end{figure}

In the left panel of Figure~\ref{fig:hhhidm1}, we depict the corrections to the triple Higgs coupling as a function of the four-momentum $q$. The color code indicates the allowed charged scalar mass. As a reference point, we display
by a solid-red line the relative corrections $\Delta\Gamma_{hhh}^{SM}$  to the trilinear Higgs boson self-coupling within the SM. 
 From this plot, it can be seen that the corrections are small and consistent with the SM predictions for light charged and neutral inert particles, whose masses are in the range $80 \textrm{ GeV} \leq m_{H^{\pm}} \leq 200 \textrm{ GeV}$ and $	100 \textrm{ GeV} \leq m_{H} \leq 200 \textrm{ GeV}$. It can also be observed that for inert scalars masses, $300 \textrm{ GeV} \leq m_{H^{\pm}}, m_{H} \leq 440 \textrm{ GeV}$, the deviation of the triple Higgs coupling from the SM's predictions is significant and
 larger than 10\% with an enhancement up to $120\%$ for $q=880$ GeV. Furthermore, One can infer that for heavy inert scalars, the corrections are substantial in  a large part of the parameter space with an enhancement of 472\% for $q=1216$ GeV and $m_H=608$ GeV. It is worth mentioning that in the left panel there are two different threshold peaks which are attributed to the opening of $h^* \to H^\pm H^\mp$ for $q=1204$ GeV with $m_{H^{\pm}}=602$ GeV, where the corrections can go up to 470\%. The second spike at $q=1216$ GeV which amplifies the radiative corrections corresponds to the threshold effect in $h^* \to HH$ with $m_H=608$ GeV. The non-decoupling effect in the radiative correction to the trilinear self-coupling of the Higgs boson is significant when large masses of the inert scalars are involved, this behavior can be seen on the right panel of Figure~\ref{fig:hhhidm1}. It is noteworthy to highlight that this behavior is also observed in the 2HDM, where large corrections to the triple Higgs coupling at the one-loop level are found to grow as the quartic power of the extra heavier Higgs bosons\cite{Kanemura:2002vm,Kanemura:2004mg,Kanemura:2017wtm}.

\section{Conclusion}
\label{sec:conclusions}
We computed the trilinear Higgs boson coupling in the IDM at one-loop level with 
non-degenerate inert scalar masses, taking into account all relevant theoretical and experimental constraints, including dark matter searches and Higgs boson invisible decay. We evaluated the one-loop Feynman amplitudes using dimensional regularization in the 't Hooft-Feynman gauge and employed an on-shell scheme renormalization. Our results showed substantial deviations from the SM predictions for the triple Higgs coupling, with values exceeding $100\%$ in some regions of the parameter space and reaching up to $472\%$
enhancement due to non-decoupling effects of the inert scalars. This substantial non-decoupling correction to the triple Higgs boson self-coupling is known to be associated with a strongly first order electroweak phase transition, which is necessary for successful electroweak baryogenesis. Detecting significant deviation from the expected
value in the triple Higgs coupling in future colliders can indirectly provide information on the mass of inert scalar bosons.

%%%%%%%%%%%%%%%%%%%%%%%%%%%
\bibliographystyle{JHEP}
\bibliography{biblio}
\end{document}